\documentclass[12pt,fullpage]{article}
\usepackage{subfigure}
\usepackage{mathrsfs}
\usepackage{amssymb,latexsym}
\usepackage{amsmath}
\usepackage{graphicx}
\usepackage{array}
\usepackage{psfrag}
\usepackage{setspace}
\begin{document}
\bibliographystyle{unsrt}
\title{Ion specific effects on phase transitions in protein solutions }
\author{S. Lettieri, Xiaofei Li and J. D. Gunton\\
\textit{Department of Physics, Lehigh University, Bethlehem, Pa 18015}} \maketitle
\begin{abstract}
A recent Monte Carlo simulation determined the potential of mean force between two lysozyme molecules in various aqueous
solutions [M. Lund et al. Phys. Rev. Lett. 100, 258105 (2008)]. The study involved a combination of explicit solvent and continuum model simulations and showed that
there are significant ion-specific protein-protein interactions due to hydrophobic patches on the protein surfaces.  In this
paper we use the results of their study to determine the phase diagram for lysozyme for aqueous solutions of NaCl and NaI.  Two of
the three phase diagrams have a stable fluid-fluid critical point, while the third has a slightly metastable critical point. This results from a secondary extremum in the potential associated with a repulsive interaction. This repulsive interaction reduces the effective range of the attractive interaction and produces a metastable critical point. We compare the results of one of these phase diagrams with that for a model that includes ion-dispersion forces, but does not
contain solvent structural effects.

\begin{singlespace}
\end{singlespace}
\end{abstract}

\newpage
\section{Introduction}
Hofmeister effects refer to the relative effectiveness of anions or cations on a wide range of phenomena and date back to the original work of Hofmeister in 1888 \cite{Hofmeister_88_01}.  Examples include the surface tension of electrolytic solutions, bubble-bubble interactions, micelle and microemulsion structure and wettability \cite{Ninham_97_01}. Hofmeister's original observation was that
protein salting out depends on salt type, as well as on the salt concentration.  In general, the effective interactions between most charged and neutral objects in aqueous solutions depend not only on the salt concentration, but also on the salt type.  This fact has remained a challenge to theorists; a wide range of explanations have been proposed \cite{Kunz_04_01,Gunton_07_01},
most of which are not of a quantitative nature.  Recently, however,  there have been two parallel developments that have provided considerable insights as to the causes of the Hofmeister effect.  The first approach emphasizes the importance of including nonelectrostatic, ion-dispersion
forces \cite{Ninham_97_01,Kunz_04_01,Bostrom_06_01,Lima_07_01} that had previously been neglected in standard treatments such as the Derjaguin-Landau-Verwey-Overbeek theory.  In this treatment water is treated as a continuum, whose properties are described by a bulk dielectric constant and ion-dispersion forces between ions and solute particles are included, together with the standard
Coulomb and van der Waals interactions to obtain a more self-consistent theory.  In the second approach molecular dynamic simulations have been carried out to obtain the effective interactions  between ions and interfaces, including the air-water interface \cite{Jungwirth_06_01,Chang_06_01} and the hydrophobic solid-water interface \cite{Marrink_01_01,Horinek_07_01}.  In
these MD simulations, water, surface and ion interactions are described by a phenomenological model (parametrized  by Lennard-Jones interaction ranges  and depths,  and partial charges) whose parameters are chosen to match experimental bulk data.
These studies have been successful in explaining a variety of ion-specific results, including explaining the adsorption of weakly hydrated ions such as bromide or iodide at the air-water interface\cite{Jungwirth_06_01,Chang_06_01} and the fact that large
halide ions are attracted to hydrophobic solid surfaces, while smaller anions are repelled\cite{Horinek_07_01}. These two complementary approaches have demonstrated the importance of including short-range interactions that account for both ion dispersion forces and short-range ion hydration effects.  Indeed, a recent paper  uses a parametrized potential of mean force that includes both ionic dispersion forces and short-range ion hydration to study ion specific effects for the double layer pressure between two uncharged interfaces\cite{Lima_08_01}.

The most recent application of molecular dynamics simulations has been to provide a step toward including a solvent induced ion-specific surface interaction in a Monte Carlo (MC) study of the interaction between lysozyme molecules in an aqueous salt solution\cite{Lund_08_02}.  In this study the solvent is treated as a dielectric continuum, but solvent structure effects are
implicitly included.  The authors obtained the potential of mean force between two lysozyme molecules in various solutions of sodium chloride and sodium iodide, respectively.  The corresponding second virial coefficients were shown to be in reasonable agreement with experimental data.  They find that there are at least two effects responsible for the Hofmeister series in this approach. Namely,
it is the direct interaction of hydrated anions with positively charged amino acid residues as well as the affinity of these anions for hydrophobic patches at protein surfaces.  The former interactions are stronger for chloride than for iodide, whereas the opposite is true for the second effect.  Thus the effective protein-protein interaction in a particular salt solution results
from a subtle balance between these (and perhaps other) effects.  The fact that there is a stronger lysozyme-lysozyme interaction in aqueous NaI than NaCl is due to the hydrophobic effect of iodide surpassing the ion-pairing effect of chloride.

An important property of an electrolytic solution of proteins is its phase diagram. It has been shown for a significant number of proteins that optimal crystal nucleation tends to occur when the solution is prepared in the vicinity of a metastable critical point of a protein-poor, protein rich fluid-fluid phase separation curve\cite{George_94_01,Rosenbaum_96_01,tenWolde_97_01}. Thus it is of interest to know the phase diagrams of the model of lysozyme obtained in the above molecular dynamic simulations.  In this paper we present MC simulations of these phase diagrams for several of the electrolyte solutions considered in reference\cite{Lund_08_02} and compare the results of one
phase diagram (0.2 M NaCl) with a corresponding calculation using a model based on ion-dispersion forces\cite{Bostrom_06_01,Lettieri_08_01}.  We present in Section 2  a summary of the model and our simulation details.  We give our results in Section 3
and present a brief conclusion in section 4.

\section{Model and Methods}
The model of  lysozyme  in aqueous salt solution model, as given by Lund et al.\cite{Lund_08_02}, consists of the following: each amino acid on the protein is represented as a soft charge-neutral sphere, located at the residue center of mass given by X-ray structure $4LZT$ in the protein data bank.  In addition,  the protonation sites of all titratable groups are included at their
original positions and their corresponding electrostatic charges are set for a pH of $4.7$.  The rigid protein molecules are allowed to rotate and translate in their MC simulations, while mobile salt and counter ions are explicitly included as soft spheres.  The solvent is treated as a dielectric continuum and solvent structural effects (hydrophobicity) are
\textbf{implicitly} accounted for. Specifically, the interaction energy between the ions and the nearest hydrophobic residues, $V_{sp}(r_{ij})$, is based on an empirical potential, from results obtained from MD simulations of ions in the presence of an air/water interface \cite{Horinek_07_01,Dang_02_01}. The expression for the total interaction energy, from which the angular averaged potential of mean force (PMF) was calculated,  is given as
\begin{equation}
\beta U = \sum{ \beta V_{sp}(r_{ij}) }+\sum{\frac{l_{B}z_{i}z_{j}}{r_{ij}}}+4\epsilon_{LJ}[(\frac{ \sigma_{ij} }{ r_{ij}
})^{12}-(\frac{ \sigma_{ij} }{ r_{ij} })^{6}]
\end{equation}
where $l_{B} = 7.1 {\AA}$ is the Bjerrum length, $r_{ij}$ is the distance between particles i and j, z is their charge number and $\epsilon_{LJ}$ and $\sigma_{ij}$ are the Lennard-Jones parameters. From the above expression for $\beta U$, Lund et al. used MC simulations to calculate the angular averaged PMF, $W(r)$, between the lysozyme molecules for various aqueous electrolyte solutions. With their PMF data, we fit these potentials for the following systems:  $0.2M$ NaCl, $0.3M$ NaCl and $0.2M$ NaI, as
shown in Fig. \ref{potentials}. It is important to note that the simulations carried out in Lund et al. were conducted at $T=298K\equiv T_0$. We use these potentials for the range of temperatures considered in this paper.  For each model, we have defined the corresponding value of $\sigma$ to be the smallest value of $r$ at which the potential crosses $W = 0$.  For the three models examined ( $0.2M$ NaCl, $0.3M$ NaCl and $0.2M$ NaI ), these values are nearly identical and are equal to $28.84 {\AA}$, $28.84 {\AA}$ and $28.88 {\AA}$ respectively.  Therefore, for the remainder of the paper, $\sigma$ is to be interpreted as the $\sigma$ for the corresponding model being referenced. In addition, the potentials have been set equal to $0$ for values of $r/\sigma$ greater than $1.8$, $2.06$ and $2.02$, respectively.

\subsection{Methods}
Using these PMFs, we perform MC simulations in order to determine the corresponding phase diagrams and pair correlation functions.  Our systems consist of  N=$500$ particles in cubic simulation cells subject to periodic boundary conditions.  The same number of MC cycles are performed for both equilibration and production, although the total number varies depending on the
type of simulation. A single MC cycle is defined as N=500 MC steps where a step is a random choice from the usual repertoire of MC moves.
\subsubsection{Fluid-Fluid coexistence}
We use the Gibbs Ensemble MC method \cite{Panagiotopoulos_87_01,Frenkel_02_01} to obtain the equilibrium coexisting densities of the protein-poor and protein-rich fluid phases.  This method avoids problems associated with the formation of an interface between the dense
and dilute fluid phases that would otherwise be present in single cell simulations. In this ensemble, two physically separate simulation cells are coupled to the same heat bath and are used to emulate the two fluid phases that are in contact.  Standard particle displacements are performed within each simulation cell; in addition, volume and particle exchanges are performed between
the two cells.  These exchanges are chosen such that the total volume and number of particles of the system are conserved and the simulations obey detailed balance. On average, we chose the ratio of particle displacements to volume moves to be 100:1; the frequency of particle transfers is chosen to give reasonable acceptance rates of approximately 1-3$\%$. The equilibrium and
production run times are at least $2*10^{5}$ MC cycles each, where a  MC cycle in this ensemble is N=500 attempts at one of three randomly selected trial moves: particle displacements, volume exchanges or particle exchanges.
\subsubsection{Fluid-Solid coexistence}
Fluid-solid coexistence are obtained via the Gibbs-Duhem method. This method involves integrating the first-order Clausius-Clapeyron equation $\frac{dP}{d\beta}=-\frac{\Delta h}{\beta\Delta v}$ where P is the system pressure, $\beta=1/kT$, $\Delta$ indicates a difference between the liquid and solid phases and $h$ and $v$ are the molar enthalpies and volumes
respectively.  One caveat to this approach is that it requires the knowledge of an initial coexistence point on the $\beta P-\beta\mu$ plane. Consequently, we carry out a series of NPT simulations along an isotherm to determine the equation of state for both the fluid and solid phases.  The equilibrium and production times for each NPT simulation are taken to be equal and at
least $2*10^{5}$ MC cycles.  Each isotherm requires a minimum of $50$ data points to obtain accurate fits.  Once the equations of state are obtained, they are fit to the following form for the liquid $\beta P_{Liq} = \sum^{m}_{i=1}c_{i}(\frac{\rho}{1-a\rho})^{i}$ and $\beta P_{Sol} = \sum^{4}_{i=0}c_{i} \rho^{i}$ for the solid, where the $a$, $c_{i}$ and $m$ are distinct for each isotherm examined and chosen to best fit the data (Whereas in other studies, one typically use m=3, we find it necessary to increase it to 5 or 6 in cases studied here.). Integration of the equations of state
will yield the free energies and chemical potentials for the liquid and solid.  However, as a result of the integration, one must know the free energy of the solid at a particular reference density along the solid isotherm. To calculate this free energy, we use the Frenkel-Ladd method for soft-core continuous potentials \cite{Ladd_84_01} to
harmonically couple the particles to lattice sites. With the free energy of the solid at the reference state known, straightforward integration produces the expressions for the corresponding chemical potentials and thus the initial coexistence points are determined via the conditions for mechanical and chemical coexistence in equilibrium, i.e. $P_{Liq}(\rho_{Liq})=P_{Sol}(\rho_{Sol})$ and $\mu_{Liq}(\rho_{Liq})=\mu_{Sol}(\rho_{Sol})$ where $\rho_{Liq}$ and $\rho_{Sol}$
are the coexisting liquid and solid densities.
\subsubsection{Accurate determination of $T_c$}
It is known that systems with short-range attractive interactions ($<1.25\sigma$) give rise to liquid-liquid phases separations that are metastable with respect to the solubility curve\cite{Pagan_05_01}, i.e., freezing preempts fluid-fluid phase separation. Choosing as an initial state a position near a metastable fluid-fluid critical point is desirable for experimentalists
trying to grow high quality protein crystals from solutions as nucleation rates reach a maximum in that region \cite{tenWolde_97_01}.  An important observed characteristic of metstable fluids is their
small and negative second virial coefficients $B_2$ first noted by George and Wilson\cite{George_94_01}.   As will be discussed in more detail in our results, the phase diagram for the 0.3M NaCl system is only very slightly metastable as a consequence of an additional repulsive maximum in the potential.  This is important because the system would otherwise exhibit a
stable fluid-fluid transition without the addition of the repulsive maximum.  Therefore in an effort to substantiate that this system is indeed metastable, an accurate estimate of the critical temperature is determined by the Bruce-Wilding finite-size scaling method \cite{Wilding_95_01}.  The location of the critical point can be obtained by matching the probability distribution of the ordering operator $M=\frac{\rho-su}{1-sr}$ with the universal distribution characterizing the Ising class. The number density and energy density are defined by $\rho=L^{-d}N$ and $u=L^{-d}U$, respectively, with $U$ the total energy of the system, $d$ the dimension of the system. To obtain the value of $T_{c}(L)$ for a given simulation cell length $L$, we run a series
of Grand Canonical Ensemble simulations and ``tune" the values of chemical potential, temperature and $s$ until the probability distribution of the ordering operator $P(M)$ collapses onto the universal $P^{*}(M)$ for the Ising model.  When good estimates of the tuning parameters are obtained, we perform longer runs of $2*10^5$ MC cycles to collect accurate statistics followed by a
histogram reweighting procedure to obtain accurate values of the tuning parameters at criticality.  This procedure is repeated for increasing values of $L$ and then extrapolated to the limit $L=\infty$ to determine the critical temperature of the infinite system.  A detailed discussion on this procedure can be found in Li et al. \cite{Li_08_01}.

\section{Results}
\begin{figure}
\includegraphics[width=3.5in]{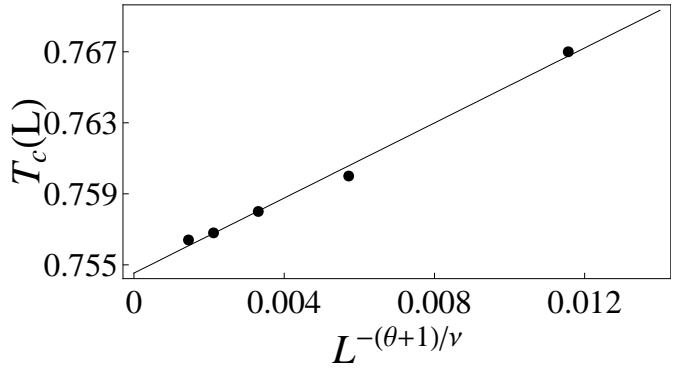}
\caption{FSS results for $T_c(L)$ as a function of $L^{-(\theta + 1)/\nu}$ for the 0.3M NaCl electrolyte solution. $\nu$ is the critical exponent for the correlation length and $\theta$ is the universal correction to scaling exponent. For the 3D Ising universality class, $\nu=0.629$ and $\beta=0.326$. By extrapolating to infinite volume ($L\rightarrow \infty$), we can obtain an estimate of the true bulk behavior: $T_c(\infty)\simeq 0.755T_0$. For more details of FSS, see Ref. \cite{Li_08_01}}
 \label{tvsl_nacl03}
\end{figure}

\begin{figure}
\includegraphics[width=3.5in,height=3.0in]{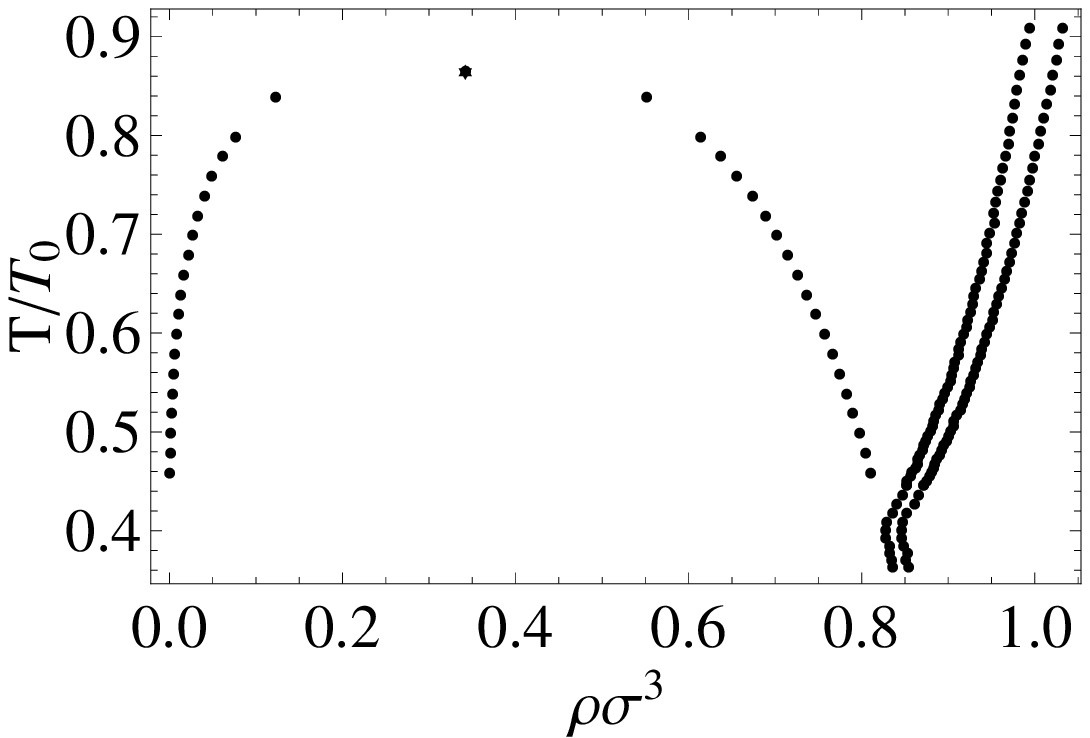}
\caption{Phase diagram curve obtained from Monte Carlo simulations for the aqueous lysozyme solution with NaCl electrolyte at 0.2M. The fluid-fluid separation curve is stable.  Solid circles are simulation data and the star is our estimate of the critical point as determined by a rectilinear diameter fit.  Note that the data for liquid line overlaps the data from the dense fluid phase. } \label{2mnaclphase}
\end{figure}

\begin{figure}
\includegraphics[width=3.5in,height=3.0in]{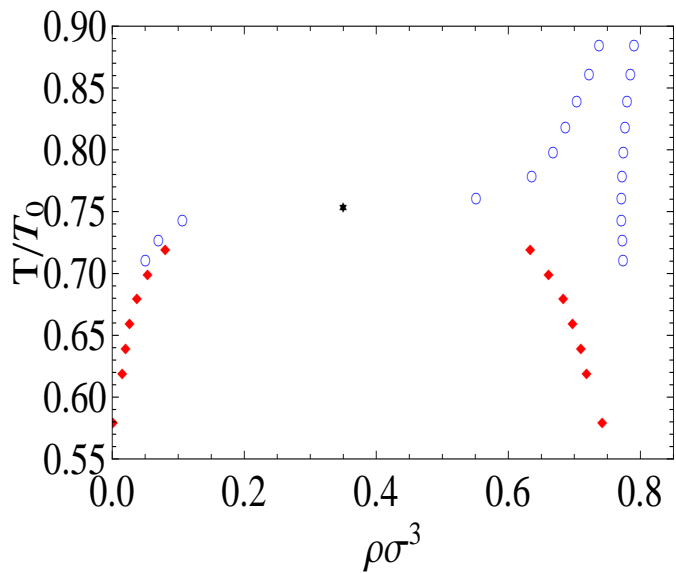}
\caption{Phase diagram curve obtained from Monte Carlo simulations for the aqueous lysozyme solution with NaCl electrolyte at 0.3M. The fluid-fluid separation curve is metstable.  Open circles are the liquid-solid coexistence data and solid diamonds are the fluid-fluid coexistence data.  The solid star is our estimate of the critical point for the infinite system as determined from finite size scaling.} \label{3mnaclphase}
\end{figure}

\begin{figure}
\includegraphics[width=3.5in,height=3.0in]{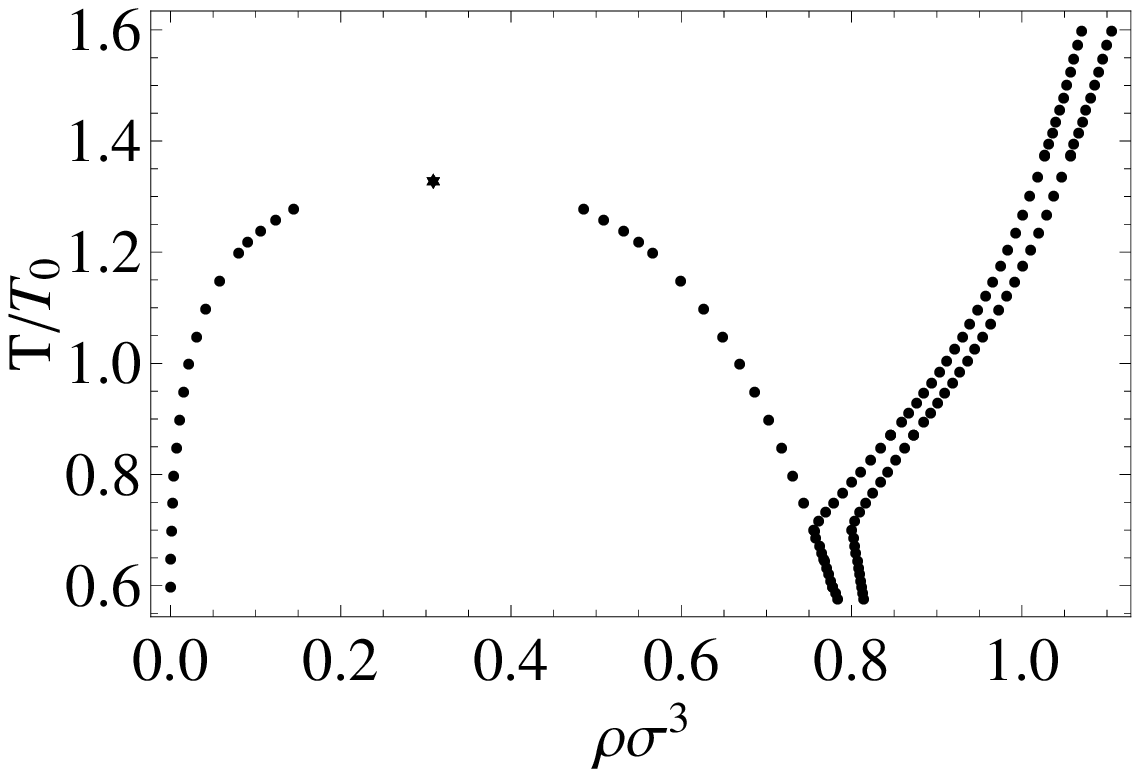}
\caption{Phase diagram curve obtained from Monte Carlo simulations for the aqueous lysozyme solution with NaI electrolyte at 0.2M. The fluid-fluid separation curve is stable.  Solid circles are simulation data and star is our estimate of the critical point as determined by a rectilinear diameter fit.  Note that the data for liquid line overlaps the data from the dense fluid phase.} \label{2mnaiphase}
\end{figure}

We have determined the complete phase diagrams for three different aqueous solutions of lysozyme, 0.2 M and 0.3 M NaCl and 0.2 M NaI shown in Figures \ref{2mnaclphase},\ref{3mnaclphase} and \ref{2mnaiphase} respectively. In the case of 0.2 M NaCl, we obtain a
``normal" phase diagram, in that the fluid-fluid coexistence curve is stable. In the case of 0.3 M NaCl,  we find that there is a metastable fluid-fluid coexistence curve, although the metastability gap (defined as $(T_L-T_c)/T_c$, where $T_c$ denotes the critical temperature for the finite system, and $T_L$ denotes the temperature on the solid curve that corresponds to the critical density) is very
small. We estimate the triple point for the 0.2M solution to be approximately 0.4 $T_0$. The situation for the 0.3M NaCl solution is a little trickier, since the solid branch of the solid-liquid  coexistence curve almost touches the fluid-fluid curve. In order to determine the critical point for the infinite system, we employ the method of finite size scaling (FSS), discussed briefly in section 2.1.3.  Our results for $T_c(L)$ are shown in Fig. \ref{tvsl_nacl03}. (See the figure caption for more details.) The extrapolation $L \rightarrow \infty$ gives us a reliable estimate of the critical point, $T_c(\infty)$,  for the bulk system.  However,  the NPT simulations we use to determine the solid-liquid curve are also subject to finite size effects and we are unable to extrapolate these to obtain their bulk limits. Therefore it is difficult to be precise about the magnitude of the metastability gap. Thus it would seem prudent to conclude that fluid-fluid critical point for 0.3M NaCl is  right on the edge of being metastable. In contrast to our results for  this model, another model that takes into account ion-dispersion forces
\cite{Lettieri_08_01} leads to a large metastability gap of $8.1\%$.

For the 0.2M NaCl and 0.2M NaI solutions,  we obtain estimates of the critical points by fitting the fluid-fluid coexistence data to the following standard equations
\begin{equation}
\frac{\rho_{l}+\rho_{g}}{2} \simeq \rho_{c}+A|T-T_{c}|\label{eq:coex1}
\end{equation}
\begin{equation}
\rho_{l}-\rho_{g} \simeq B|T-T_{c}|^{\beta}\label{eq:coex2}
\end{equation}
where $T_{c}$ and $\rho_{c}$ are the critical temperature and density, respectively, and $\beta$ $\simeq$ 0.326 is the 3D-Ising critical exponent. Finite size effects would have to be taken into account to obtain the critical point parameters for the infinite system. In the case of 0.2M NaI ( Fig. \ref{2mnaiphase} ), we estimate the triple point temperature to be around 0.7 $T_0$. We are not able to make direct comparisons between simulation results and experimental data for either salt, however, since to the best of our knowledge, no experimental data exist for the molarities studied in this paper. However, for solutions with 0.5M or
greater concentrations of NaCl, experimental data is available. This data shows the presence of a metastable fluid-fluid phase for all salt concentrations greater than 0.5M \cite{Muschol_97_01}. It is also likely that there is a metastable critical point for concentrations somewhat smaller than 0.5M, given that the NaCl solution at that molarity exhibits a large metastability gap\cite{Muschol_97_01}.

It is also interesting to note that for both stable phase diagrams ( Figs. \ref{2mnaclphase} and \ref{2mnaiphase} ), there is a metastable continuation of the liquidus curve along the high density branch of the fluid-fluid coexistence curve below the triple point, $T_{tp}$. Namely, as we continue our simulation of the coexisting liquid phase in the liquidus line below $T_{tp}$, this becomes the metastable dense liquid phase we have obtained by our gibbs ensemble method.

\begin{figure}[]
\begin{center}
    \mbox{

     \subfigure[]{\scalebox{0.6}{\includegraphics[width=4.5in,height=3.5in]{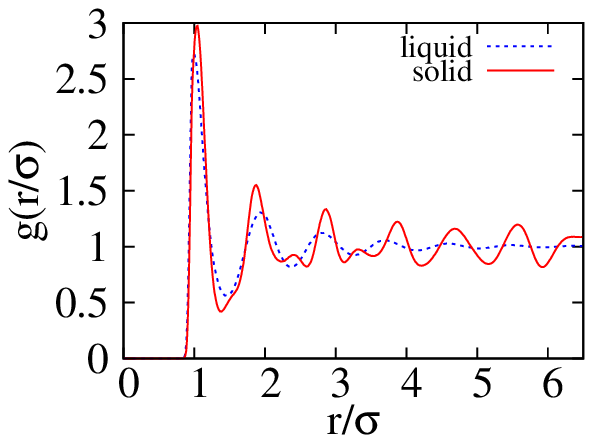}}} \quad
      \subfigure[]{\scalebox{0.6}{\includegraphics[width=4.5in,height=3.5in]{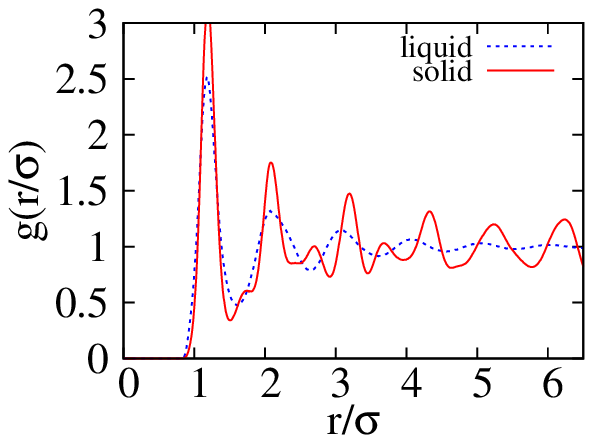}}}
          }
       \caption{Plot of the liquid and solid pair correlation functions corresponding to coexisting liquid and solid states for (a) 0.2M NaI at $T=1.6 T_0, P=20.409 \sigma^3/kT_0,  \rho_{Liq}\sigma^{3} =1.069 , \rho_{Sol}\sigma^{3}=1.104$ and (b) 0.3M NaCl at $T=0.91 T_0, P=2.002\sigma^3/kT_0, \rho_{Liq}\sigma^{3} = 0.754, \rho_{Sol}\sigma^{3} =0.798 $.}
\label{correlations}
\end{center}
\end{figure}

Figure \ref{correlations} plots the pair correlation functions of the liquid and solid at conditions corresponding to phase coexistence for the 0.2M NaI and 0.3M NaCl lysozyme solutions, respectively.  In order to show that the solid does indeed exhibit long range order, we ran simulations of N=4000 particles to probe larger distances.  It is clear that both liquid and solid correlation functions display an oscillatory behavior. The reason that it does so for the liquid phase is due to its high density. Note that in Figures \ref{2mnaclphase} and \ref{2mnaiphase} the separation
between the  liquid and solid coexistence densities is very small. However, the correlation functions for the solid still shows a strong peak at the largest distances, suggesting a crystal structure, while it already decays to 1 for the liquid phase after a few ($\sim4.5$) $\sigma$s.

\begin{figure}[]
\begin{center}
    \mbox{

     \subfigure[]{\scalebox{0.6}{\includegraphics[width=3in,height=2.3in]{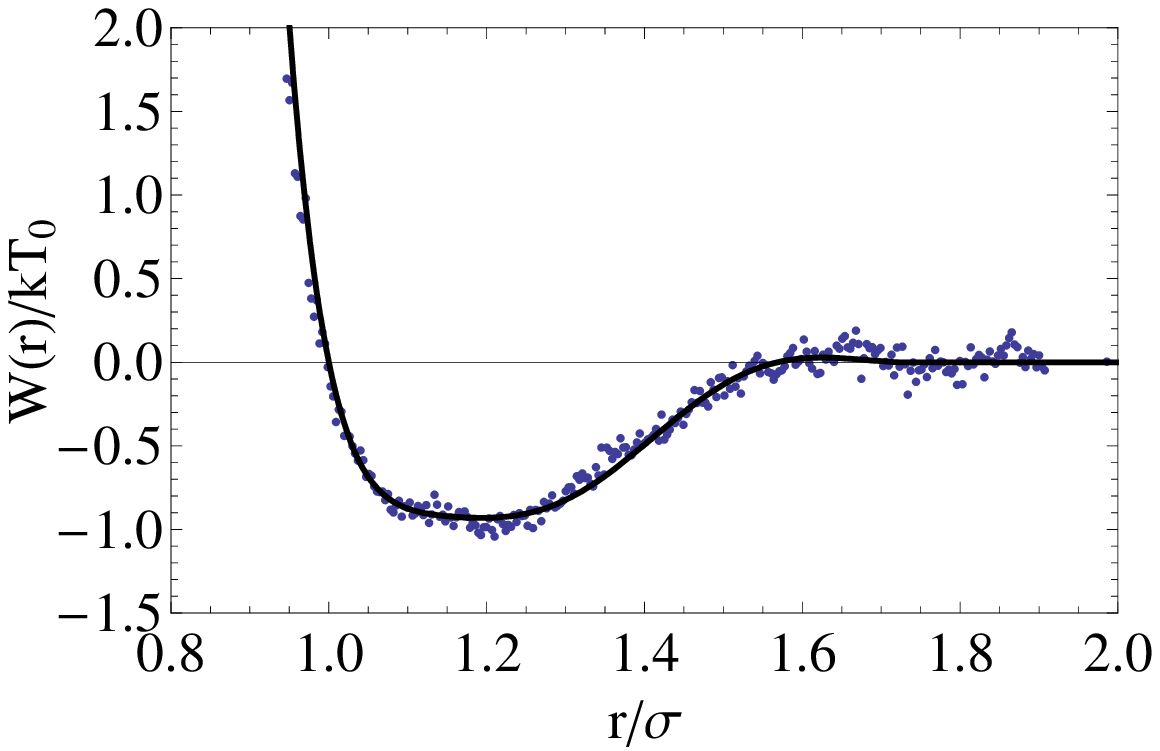}}} \quad
      \subfigure[]{\scalebox{0.6}{\includegraphics[width=3in,height=2.3in]{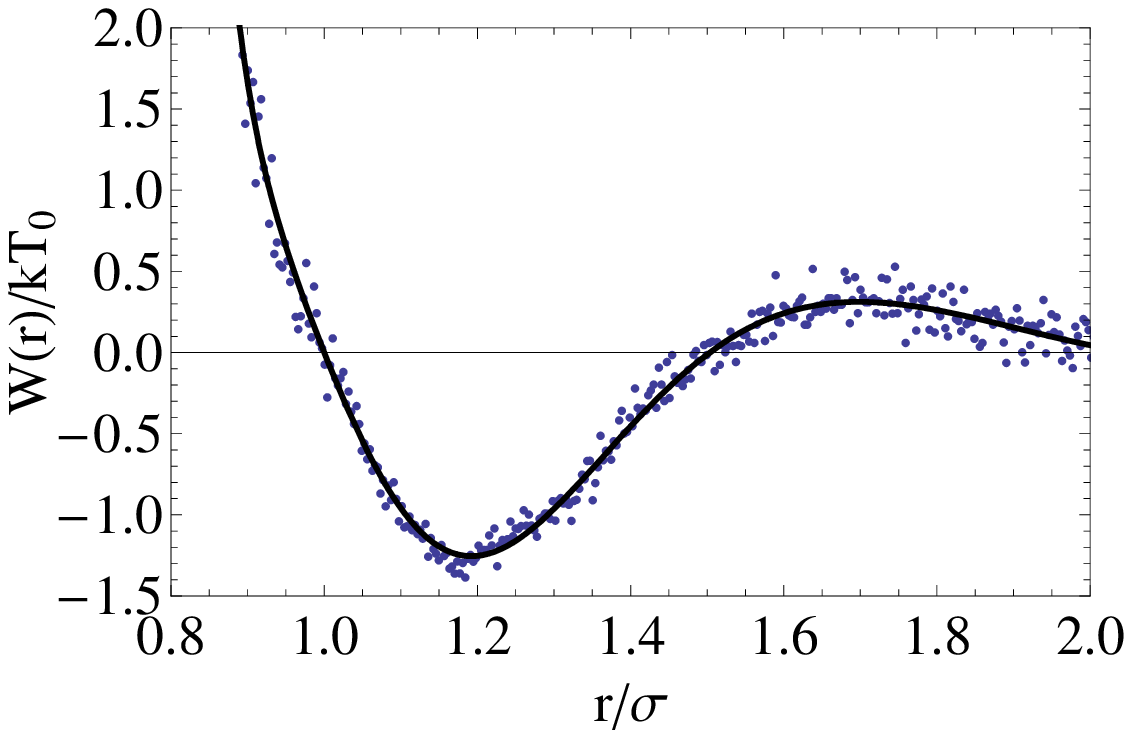}}}\quad
      \subfigure[]{\scalebox{0.6}{\includegraphics[width=3in,height=2.3in]{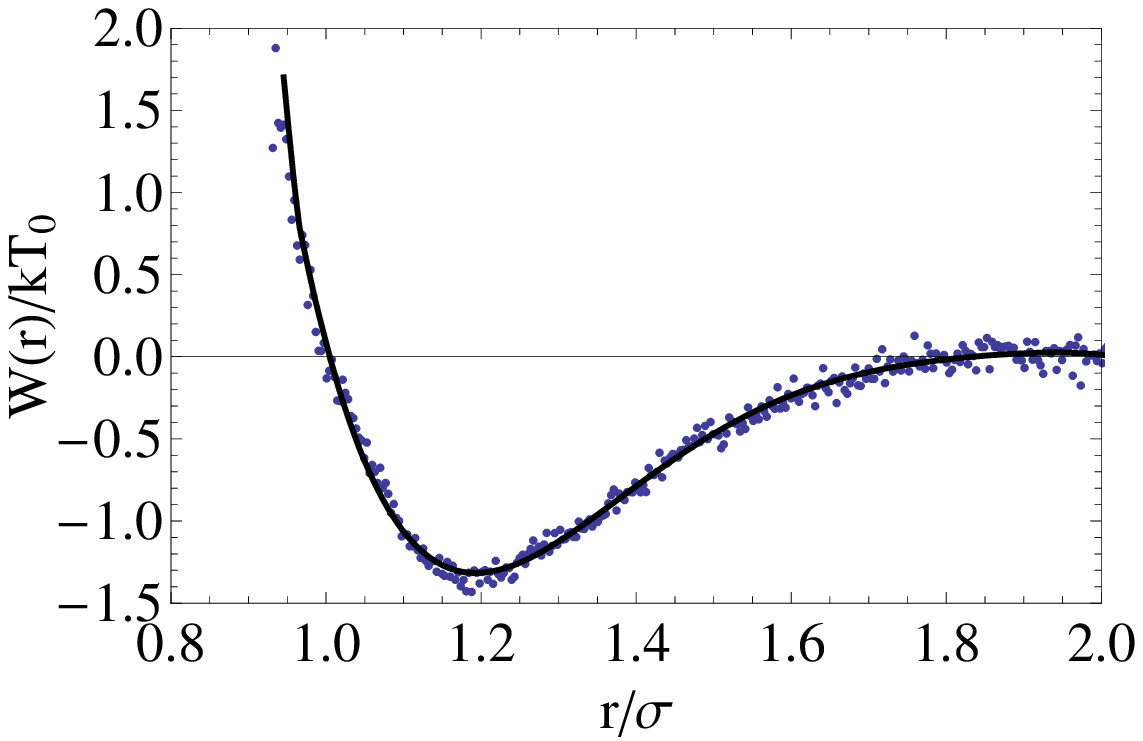}}}}

       \caption{Total potentials for the (a) 0.2M NaCl (b) 0.3M NaCl (c) 0.2M NaI electrolyte systems.
   Solid line is the fitted polynomial and dots are data obtained from MC simulations.}
\label{potentials}
\end{center}
\end{figure}

One can see from the phase diagrams of Figs. \ref{3mnaclphase} and \ref{2mnaiphase} that the details of the corresponding potentials ( Figs. \ref{potentials}(b) and \ref{potentials}(c) ) are important in determining the phase diagram. Although the attractive wells of their potentials are similar in both the width\cite{Footnote_08_01} and the depth, their phase diagrams are very much different. The 0.3M NaCl solution has a metastable
fluid-fluid coexistence curve while the 0.2M NaI solution has a stable phase separation curve. This can be largely attributed to  the fact that the potential for the 0.3M NaCl has a repulsive region which reduces the effective range of the attractive interaction.  In other words, the critical point of the fluid-fluid coexistence curve is driven to a lower temperature by the additional positive extremum in Fig. \ref{potentials}(b). This behavior was also observed
by Brandon et al. \cite{Brandon_06_01}, in their study of models with multiple extrema.  They argued that for
more complicated potentials than those with a single extremum, the effective range of attraction defined by Noro and Frenkel\cite{Noro_00_01} was more useful in characterizing the phase diagrams than, say, the width of the attractive well.  Noro and Frenkel\cite{Noro_00_01}  developed an  extended law of corresponding states, in which they introduced an energy scale (the depth of the attractive well), an effective hard core $\sigma_{eff}$ and a
range $R$.  To do this  for systems whose potentials are continuous, one divides the potential into attractive and repulsive terms, $v_{att}$ and $v_{rep}$, respectively, and defines an ``equivalent" hard-core diameter for the repulsive part of the potential using an expression suggested by Barker and Henderson\cite{Barker_76_01}:
\begin{equation}
\sigma_{eff} = \int_{0}^{\infty}dr[1-\exp(-v_{rep}/kT)].
\end{equation}
The reduced second virial coefficient is then defined as $B_{2}^{*}=B_2/(\frac{2\pi\sigma_{eff}^{3}}{3})$.    One then defines an effective range $R$ of the attractive interaction by  equating $B_{2}^{*}$ of a square well system with that of the system in question. (Note that the range of the attractive interaction for a square well system is defined unambiguously, whereas this is not the case for other models.) Although this range is temperature dependent, it provides a useful length scale for models such as those
that have been used in colloidal and protein solutions. The value of this reduced second virial coefficient at the critical point has been found to be relatively constant for a wide range of models that are commonly used to describe phase transitions, ranging from  extremely narrow attractive wells (Baxter's adhesive hard sphere model) to the van der Waals limit of infinitely long-range attractive wells\cite{Noro_00_01}.  These values are in the range $-2.36 \leq B_{2}^{*}(T_c)\leq -1.33$.  We show the reduced second virial coefficient $B_{2}^{*}(T_c) $, the effective hard core diameter $\sigma_{eff}$  and the range $R$ for all three potentials  studied here in Table \ref{b2table}.   All our values for $B_{2}^{*}(T_c)$ are in the
same range as for the models cited in Ref. \cite{Noro_00_01}. Noro and Frenkel also estimated that the fluid-fluid transition became metastable with respect to freezing when $R = 0.14$, consistent with a variety of models that had been studied at that time. Subsequently, however, it was shown that for the square well model $R\simeq 0.25$\cite{Pagan_05_01}.  This latter value has been proposed based on a simple van der Waals model for both the fluid and the solid phase\cite{Daanoun_94_01} and by a phenomenological argument using a cell model for the solid\cite{Asherie_96_01}.  If we include the region of repulsive interaction associated with the secondary maximum in our calculation of the effective hard core diameter, we find that the range for the 0.3 M NaCl system that is slightly metastable is $R\simeq 0.20$.

Finally, we note that we have also run preliminary NPT simulations for the 0.3M NaI solution.  However, no phase diagram was calculated due to the fact that the difference in density between the liquid and solid isotherms is extremely small even when the pressure is sufficiently high.  In addition, at higher pressures, the density on the liquid isotherm has large fluctuations about its average; in fact, it typically transitions from the liquid phase to the solid phase where it remains stuck for the remainder of the simulation.  Therefore, since we are unable to distinguish sufficiently accurately between the liquid and solid phases, we are unable to perform an accurate calculation of the 0.3M NaI phase diagram by this method.
\section{Conclusion}
We have studied models of three electrolyte solutions of lysozyme and demonstrated the effects of hydrophobic surface forces on the phase diagram of lysozyme.  Although the molarities studied are too small to see a significant metastability gap, the 0.3 M NaCl solution does have  a slightly metastable fluid-fluid critical point.  Interestingly, the small range $R$ associated with this is produced by the effect of a repulsive interaction associated with a secondary extremum.  In the absence of that repulsive region the system would have a stable fluid-fluid transition.  This is another example of a phenomenon first studied by Brandon et al. \cite{Brandon_06_01}.  It is not clear what the physical origin of the secondary maximum is for the model of Lund et al. \cite{Lund_08_02}, but Brandon et al. \cite{Brandon_06_01} attribute their secondary maximum to the effect of water restructuring near the solute particles.  It would be interesting to have a better understanding of its origin for the current model.  As we are unable to compare the results of this study with any experimental results for lysozyme, we cannot determine the accuracy of the model.  It would seem likely that further developments will include in addition the specific ion-dispersion effects discussed by Bostr\"{o}m et al. \cite{Bostrom_06_01}.  These will require more accurate quantum mechanical calculations of the amplitudes of the ion-dispersion forces; such calculations are currently being carried out\cite{Parsons_08_01,Parsons_08_02}.

\section{Acknowledgement} This work was funded in part by a grant from the National Science Foundation, DMR-0702890.

\begin{table}[ht]
\caption{Parameters of interest at $T_c$.} \centering
\begin{tabular}{c c c c c}
\hline \hline System & $T_{c}$ & $\sigma_{eff} $ & $B_{2}^{*}(T=T_{c})$ & $R$ \\ [0.5ex] \hline
0.2M NaCl & 0.866 $T_{0}$ & 0.976  & -2.10 & 0.377 \\
0.2M NaI & 1.33 $T_{0}$ & 0.961 & -2.18 & 0.423 \\
0.3M NaCl & 0.757 $T_{0}$ & 1.069 & -2.19 & 0.205 \\
\hline
\label{b2table}
\end{tabular}
\end{table}


\begin{thebibliography}{1}
\bibitem{Footnote_08_01} The width can be defined as the difference between the distances at which the potential equals half the attraction depth.
\end{thebibliography}
\end{document}